\documentclass[twocolumn,showpacs,preprintnumbers,amsmath,amssymb]{revtex4}
\usepackage{graphicx}
\usepackage{dcolumn}
\usepackage{bm}
\usepackage[dvips]{epsfig}
\usepackage{amsmath}
\usepackage{amssymb}
\def\bea {\begin{eqnarray}}
\def\eea {\end{eqnarray}}

\def\be {\begin{equation}}
\def\ee {\end{equation}}

\begin{document}
\title{Evidence for the reduction of nuclear level density away from the $\beta $-stability line}
\vspace{.5truecm}
\vspace{.5truecm}

\author{Pratap Roy$^{1,2}$,\footnote{Email: roypratap@vecc.gov.in} K. Banerjee$^{1,2}$, T. K. Rana$^{1}$, S. Kundu$^{1,2}$, S. Manna$^{1,2}$, A. Sen$^{1}$,  D. Mondal$^1$, J. Sadhukhan$^{1,2}$, M. T. Senthil Kannan$^3$, T. K. Ghosh$^{1,2}$, S. Mukhopadhyay$^{1,2}$, Deepak Pandit$^{1}$, G. Mukherjee$^{1,2}$, ,  S. Pal$^1$, D. Paul$^{1,2}$, K. Atreya$^{1,2}$, and C. Bhattacharya$^{1,2}$ } 
\vspace{.5truecm}
\vspace{.5truecm}
\affiliation{$^1${\em Variable Energy Cyclotron Centre, 1/AF, Bidhan  Nagar, Kolkata~-~700064, India}\\ 
$^2${\em Homi Bhabha National Institute, Training School Complex, Anushaktinagar, Mumbai~-~400094, India}\\
$^3${\em 7, Ambiganagar I street, Thudiyalur, Coimbatore -641034, Tamil Nadu, India.}}

\begin{abstract}
The isospin dependence of nuclear level density has been investigated by analyzing the spectra of evaporated neutrons from excited $^{116}$Sn and $^{116}$Te nuclei. These nuclei are populated via $p$~+~$^{115}$In and $^{4}$He~+~$^{112}$Sn reactions in the excitation energy range of 18~-~26 MeV. Because of low excitation energy, the neutron spectra are predominantly contributed by the first-chance decay leading to the $\beta$-stable $^{115}$Sn and neutron-deficient $^{115}$Te as residues for the two cases. Theoretical analysis of the experimental spectra have been performed within the Hauser-Feshbach formalism by employing different models of the level density parameter. It is observed that the data could only be explained by the level density parameter that decreases monotonically when the proton number deviates from the $\beta$-stable value. This is also confirmed by performing a microscopic shell-model calculation with the Wood-Saxon mean field. The results have strong implication on the estimation of the level density of unstable nuclei, and calculation of astrophysical reaction rates relevant to $r$- and $rp$-processes.      
\end{abstract}


\maketitle
One of the primary objectives of current nuclear physics research is to reproduce the observed abundances of different elements in the universe, and to understand the underlying physical processes behind the synthesis of these elements at different astrophysical sites. Although it is reasonably well understood how the elements up to Fe
(Z=26) are produced within the stars through fusion reactions, the understanding of the synthesis of heavier elements (Z~$>$~26) has been evasive. The mysteries regarding the astrophysical sites, as well as the nuclear data needed to describe the heavy-element nucleosynthesis, persist even after intense investigations over the past decades~\cite{Goriely,Arnould,Meyer,Busso}. Many of the astrophysical reactions related to the nucleosynthesis of heavy elements involve either neutron-rich ($r$-process) or proton-rich ($rp$-process) nuclei, making it difficult to carry out direct experimental measurements for determining the reaction rates. Since most of the relevant cross-sections are not available experimentally, the usual approach is to calculate them within the statistical Hauser-Feshbach (HF) framework~\cite{Hauser}. One of the most critical inputs in the HF calculations of nuclear reaction cross-sections is nuclear level density (NLD). Experimental information on NLD is available mostly for isotopes near stability, and for most of the unstable nuclei, only theoretical estimates are available, which tend to be highly uncertain. Therefore, experimental data to understand the variation of level density from stable nuclei to the nuclei away from the stability line is of high importance.\\
\indent The simplest and most widely used description of level density is given in terms of the non-interacting Fermi gas model (FGM)~\cite{Bethe}, 
\begin{equation}
\rho (E)= \frac {1}{12\sqrt{2}\sigma }\frac{\exp{2\sqrt{a(E -\Delta )}}}{a^{1/4}(E -\Delta )^{5/4}},
\label{eq:eq1}
\end{equation}
where $E$ is the excitation energy, $\Delta $ is the pairing energy shift~\cite{Dlig}, $\sigma $ is the spin cut-off factor, and $a$ is the level density parameter (LDP) which is directly related to the density of single-particle states near the Fermi surface. Several important phenomenological refinements to Eq.~\ref{eq:eq1} have been introduced in due course to include the angular momentum~\cite{Ericson}, shell~\cite{Igna}, and collective effects~\cite{Capote,Ilji,Jung}. On the other hand, another crucial factor, the iso-spin effects in NLD is somewhat neglected largely because these effects are expected to be small for nuclei at the valley of stability. However, the iso-spin effects in NLD can be profound for unstable nuclei which, in turn, can significantly influence the calculated astrophysical reaction rates.\\  
\indent The Fermi gas model gives a smooth dependence of the level density parameter on the mass number ($A$) which can be expressed as
\begin{equation}
a= \alpha A
\label{eq:eq2}
\end{equation}
where the proportionality constant $\alpha $ is taken either from global systematics~\cite{RIPL,Egidy1,Egidy2,Egidy3} or adjusted to match the experimental data~\cite{Pratap1,Pratap2,Pratap3,KB1,Gohil,KB2,DP}. The iso-spin, as well as the single-particle binding energy can give a dependence of the level density parameter $a$ on neutron ($N$), and proton ($Z$) numbers rather than merely on the mass number $A$. Two alternative forms of Eq.~\ref{eq:eq2} that include the $N$, $Z$ dependence have been suggested by Al-Quraishi {\it et al.}~\cite{Qura1,Qura2}
\begin{equation}
a= \frac{\alpha A}{\exp{[\beta (N-Z)^2]}}
\label{eq:eq3}
\end{equation}
\begin{equation}
a= \frac{\alpha A}{\exp{[\gamma  (Z-Z_0)^2]}}
\label{eq:eq4}
\end{equation}
where $\alpha $, $\beta $ and $\gamma $ are empirical constants. The presence of the ($N-Z$) factor in Eq.~\ref{eq:eq3} causes the level density to be maximum for $N=Z=A/2$ for a given $A$ and to decrease as the neutron-proton asymmetry (iso-spin) increases. On the other hand, the ($Z-Z_0$) factor in Eq.~\ref{eq:eq4}, where $Z_0$ is the atomic number of the $\beta$-stable isotope for the mass number $A$, reduces the level density as the nucleus moves away from the $\beta$-stability line. The expression for $Z_0$ is obtained from the fit of the semiempirical mass formula. The arguments for the reduction of the level density parameter away from the stability line can be found in Refs.~\cite{Widen,Grim1,Enge,Mustafa,Grim2,Grim3}. For the lighter mass nuclei (A~$\leq $40) where $Z_0\simeq Z\simeq N\simeq  A/2$ the different forms of $a$ give similar results, and the difference is expected to become prominent only beyond $A=40$.\\ 
\indent In Refs.~\cite{Qura1,Qura2} the validity of the different expressions for the level density parameter was tested for nuclei in the mass range of 20~$\leq$~A$\leq$~110. The state densities were extracted from the experimentally measured discrete levels, and  fitted with the Fermi gas expression using different parametrizations of $a$ as given in Eqs.~\ref{eq:eq2}~-~\ref{eq:eq4}. The analysis suggested that the ($Z-Z_0$) form provides somewhat better reproduction of the experimental data compared to the other formulations. However, the distinction was not overwhelming mainly because the complete level schemes those were used to test the models were limited to low energies (3~-~4 MeV for $A\geq 40$), and available mostly for nuclei with $\mid Z-Z_0\mid\lesssim  1$. \\
\indent Tentative supports in favor of the reduction of the level density parameter in accordance with the $Z-Z_0$ form  have also been found in some of the low energy particle evaporation studies~\cite{Rich,Zhura,Voi}. However, no such evidence was observed by Charity {\it et al.}~\cite{Char1} in the high energy fusion evaporation measurement around $A$~$\approx  $~160. The theoretical study of Charity and Sobotka~\cite{Char2} also suggests very little dependence of the level density parameter on the neutron or proton richness of the nucleus. Therefore, it is evident that the issue of the variation of the level density on the neutron-proton asymmetry is far from being resolved, and experimental data for nuclei at least two units away from the stability line {\it i.e.} $\mid Z-Z_0\mid\gtrsim  2$, is highly demanding for carrying out further tests of the different $N$, $Z$ dependent parametrization of $a$. \\ 
\indent The analysis of the evaporation spectra of light particles can be used as an excellent tool for carrying out such tests in a wide excitation energy and angular momentum range. The light-ion induced reactions are particularly advantageous in populating low excitation energies and restricting the number of effective decay channels compared to the heavy-ion (HI) route~\cite{Pratap1,Pratap2,KB1}.\\
\indent In this Rapid Communication, we report the experimental study on the $N$, $Z$ dependence of NLD investigated using the neutron evaporation spectra from $^{116}$Sn, and $^{116}$Te compound nuclei (CN) populated through the $p$~+~$^{115}$In, and ${^4}$He~+~$^{112}$Sn reactions, respectively. The CN were populated at low excitation energies so that the neutron spectra are dominated by the first-chance neutron emission leading to the residual $^{115}$Sn, $^{115}$Te nuclei in the two cases. The analysis of the neutron spectra will allow us to simultaneously investigate the level densities of the two $A=115$ isobars, the $\beta $-stable $^{115}$Sn ($Z\approx Z_0 $) and the neutron-deficient $^{115}$Te ($\mid Z-Z_0 \mid >  $2) which will provide crucial information in understanding the variation of NLD as a function of $N$ and $Z$.\\
\\
The experiment was performed using 9 and 12 MeV proton and 28 MeV $^{4}$He beams from the K130 cyclotron at VECC, Kolkata. Self-supporting foils of $^{115}$In (thickness $\approx $1 mg/cm$^2$), and isotopically-enriched (99.6$\%$) $^{112}$Sn~\cite{Hari} (thickness $\approx $2.26 mg/cm$^2$) were used as targets. The compound nucleus $^{116}$Sn ($p$~+~$^{115}$In) was populated at two excitation energies $E^\ast_{CN} =$18.2 and 21.2 MeV, whereas $^{116}$Te ($^4$He~+~$^{112}$Sn) was populated at $E^\ast_{CN} =$26 MeV. The neutrons emitted during the compound nuclear decay process were detected using six cylindrical liquid scintillator based neutron detectors (of 5-inch length and 5-inch diameter) placed at the laboratory angles of 55$^\circ $, 85$^\circ $, 105$^\circ $, 120$^\circ $, 140$^\circ $, 155$^\circ $ at a distance of 1.5 m from the target. The neutron kinetic energies were measured using the time-of-flight (TOF) technique. The start trigger for the TOF measurement was generated using the prompt $\gamma $-rays detected by a 50-element BaF$_2$ detector array~\cite{Deepak}, placed near the target position. The prompt $\gamma$-$\gamma $-peak in the TOF spectrum was taken as the time reference. The efficiencies of the neutron detectors were measured in the in-beam condition using a $\approx $35 $\mu $Ci $^{252}$Cf source~\cite{Pratap4}. Neutron-$\gamma $ discrimination was achieved by both the TOF and pulse shape measurements (PSD)~\cite{Kaushik}. The scattered neutron contributions in the measured neutron spectra were estimated and subtracted using the ``shadow bar" technique~\cite{Pratap2}. Further details on the experimental setup and data analysis techniques are available in Refs.~\cite{PR,KB}\\  

\begin{figure}
\begin{center}
\includegraphics[scale=0.58]{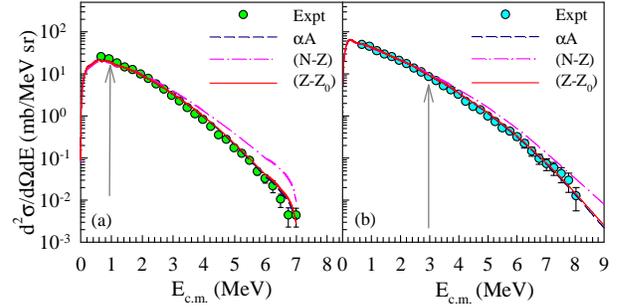}
\caption{\label{fig:fig1} Experimental neutron double differential spectra (filled circles) for the $p$~+~$^{115}$In reaction measured at 155$^\circ $ for the incident proton energies of (a) 9 MeV (b) 12 MeV. Lines are the predictions of TALYS using three different parametrizations of the level density parameter (see text). The arrows show the positions above which the spectra are entirely determined by the first-chance emission.}
\end{center}
\end{figure}

The background-corrected neutron spectra measured at various laboratory angles were transformed into the compound nucleus center-of-mass (c.m.) frame using the standard Jacobian transformation. The spectral shapes at the backward angles were found to be almost overlapping indicating the dominance of the compound nuclear contribution in the measured spectra. The spectra measured at the most backward angle (155$^\circ $) have been considered for the statistical model analysis and for testing different models of the level density parameter. The theoretical calculation of the neutron energy spectra was performed with the TALYS (v 1.9) code~\cite{TALYS} using the statistical HF framework. For the level density, the composite Gilbert-Cameron (GC) formulation~\cite{GC} was used. In the GC model, the level density at low energies (from 0 to a matching energy $E_M$) is approximated by a constant-temperature (CT) formula     
\begin{equation}
\rho_{CT}(E)= \frac{1}{T_0}exp{\frac{E-E_0}{T_0}}
\label{eq:eq5}
\end{equation}
and for energies higher than $E_M$ the level density is given by the Fermi gas expression (Eq.~\ref{eq:eq1}). The constant temperature ($T_0$), and the energy shift ($E_0$) are chosen such that the two prescriptions match together smoothly at the matching energy which varies inversely with the mass number ($A$), and for the present case $E_M\approx  $4 MeV. \\
The shell effect in NLD has been incorporated using an energy and shell-correction dependent parametrization of the level density parameter~\cite{Igna}
\begin{equation}
a(U)=\tilde{a}[1+\frac{\Delta S}{U}\{1-\exp(-\gamma U)\}]
\label{eq:eq6}
\end{equation}
where, $U=E-\Delta $, and $\tilde{a}$ is the asymptotic value of the level density parameter obtained in the absence of any shell effect. Here $\Delta S$ is the ground state shell correction,  and $\gamma$ determines the rate at which the shell effect is depleted with the increase in excitation energy~\cite{TALYS}. The transmission coefficients were calculated using the optical model (OM) where the OM parameters for neutron and proton were taken from the local and global parameterizations of Koning and Delaroche~\cite{Koning}. For the $\alpha $-particles, simplifications of the folding approach of S. Watanabe~\cite{Watanabe} is used in the TALYS calculations. It was found that the variation of the optical model parameters have very little effect in determining the spectral shape which is mainly decided by the value of the level density parameter.\\
\indent Fig.~\ref{fig:fig1} shows the experimental neutron spectra for the $p$~+~$^{115}$In reaction at the two incident proton energies. It can be seen that the experimental data are nicely reproduced by the TALYS calculation (dashed blue line in Fig.~\ref{fig:fig1}) using the standard form of the level density parameter given by the expression $\tilde{a}=\alpha A$ where the proportionality constant $\alpha$ has been taken from global systematics ($\alpha=0.1273$)~\cite{TALYS}. It should be noted that TALYS uses a non-linear dependence of the level density parameter on mass number ($\tilde{a} =\alpha_1 A+\alpha_2A^{2/3}$ with $\alpha_1 $~=0.69 and $\alpha_2 $~=0.28 for the GC model). Since the level of non-linearity is not very large and the non-linear form does not make much difference in terms of explaining the experimental data we have used the linear form for simplicity. A linearization of the non-linear expression has been done to obtain the same resultant level density parameter for the given mass number. \\
\indent Different $N$, $Z$ dependent expressions for the level density parameter as described in Eq.~\ref{eq:eq3}, and Eq.~\ref{eq:eq4} have also been tested. The values of the parameters $\beta ,\gamma $, and $Z_0$ were taken from Ref.~\cite{Qura2}. It should be mentioned here that unlike Ref.~\cite{Qura2} we have used same $\alpha$-values in the three different parametrizations of $a$. Further, the $\alpha$-value used in the present work is slightly higher compared to the values used in Ref.~\cite{Qura2} because of the difference in choosing the energy-shift $\Delta $. For the Gilbert-Cameron prescription the energy back-shift is chosen as $\Delta =\chi (12/\sqrt{A})$. Where $\chi $= 0, 1 and 2 for odd-odd, even-odd and even-even nuclei, respectively. 
For the $p$~+~$^{115}$In reaction, the neutron spectra at both the excitation energies are predominantly determined by the first-chance (1$n$) neutron emission leading to $^{115}$Sn as the evaporation residue (ER). The position of the $E_{c.m.}$ beyond which the spectra are completely decided by 1$n$ emission have been indicated by arrows in Fig.~\ref{fig:fig1}; for energies below this point there are small ($\approx  $10$\%$) contributions from the 2$n$ channel. For the $\beta $-stable $^{115}$Sn, $Z\approx Z_0$ and therefore, the ($Z-Z_0$) form provide similar results to that of $\alpha A$ form as can be seen from Fig.~\ref{fig:fig1}. On the other hand, the ($N-Z$) expression of the level density parameter could not reproduce the experimental data (dashed-dot line in Fig.~\ref{fig:fig1}) in this case. \\  
\begin{figure}
\begin{center}
\includegraphics[scale=0.40]{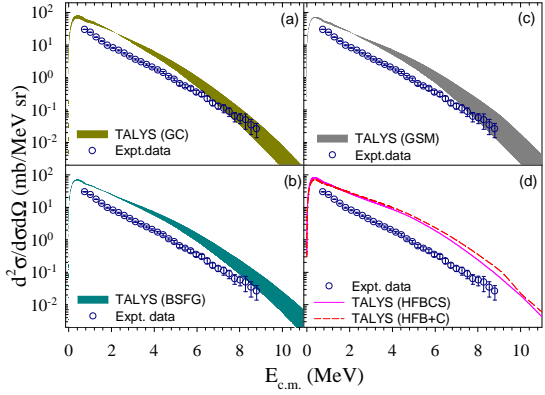}
\caption{\label{fig:fig2} Experimental neutron spectra (filled circles) for the $^4$He~+~$^{112}$Sn reaction at 155$^\circ $ along with the predictions of TALYS with phenomenological (a) GC, (b) BSFG, and (c) GSM level densities using the standard $\alpha A$ form of the level density parameter. The shaded regions in plots (a)~-~(c) corresponds to $\pm $15$\%$ variation in the standard value of $\alpha$ for each model. (d) The experimental spectrum compared with the TALYS calculation using microscopic HFBCS (continuous line), and HFB+C (dashed line) level densities as inputs.}
\end{center}
\end{figure}
\indent In contrast to the $p$~+~$^{115}$In reaction, the standard $\alpha A$ form of the LDP could not reproduce the experimental data in case of the $^4$He~+~$^{112}$Sn reaction, as shown in Fig~2. The situation could not be improved by using different forms of the phenomenological level density formulations such as the Back-shifted Fermi Gas (BSFG)~\cite{Dlig}, and the Generalized Super-fluid Model (GSM)~\cite{GSM1,GSM2} as shown in Fig~2 (b) and (c). Efforts were made to fit the data by tuning the  proportionality constant $\alpha$. However, the data could not be explained by any reasonable variation of $\alpha$ irrespective of the choice of the particular phenomenological NLD model (Fig.~\ref{fig:fig2}(a)-(c)). The shaded regions in Fig.~\ref{fig:fig2}(a)-(c) correspond to $\pm $15$\%$ variation in $\alpha $ around the default (systematic) value for each NLD model provided in TALYS. The situation could not be improved even with higher variations in $\alpha $. The microscopic level density inputs obtained under the Hartree-Fock BCS (HFBCS)~\cite{Goriely1}, and Hartree-Fock-Bogolyubov plus combinatorial (HFB+C)~\cite{Goriely2} methods also failed to reproduce the experimental data (Fig.~\ref{fig:fig2}(d)). \\
\indent Interestingly, for the $^4$He~+~$^{112}$Sn reaction, it is observed that the experimental data could be nicely explained by using reduced level density parameters given by the ($Z-Z_0$) form as shown by the continuous red line in Fig.~\ref{fig:fig3}. For this reaction at the present excitation energy (26 MeV), the most significant contributions to the neutron spectrum arise from the 1$n$ and $pn$ neutron channels leading to $^{115}$Te and $^{114}$Sb as residual nuclei, respectively. Besides, there are some small contribution ($<10\%$) from the 2$n$ and $\alpha n$ channels below $E_{c.m.}$~=5 MeV. For the most significant ERs ({\it i.e.} $^{115}$Te and $^{114}$Sb) in the $^4$He~+~$^{112}$Sn reaction $Z-Z_0\gtrsim2$. Therefore, the ($Z-Z_0$) term has a strong effect, and there is a significant reduction of the resultant level density parameters as predicted by Eq.~\ref{eq:eq4}. The experimental results as evident from Fig.~\ref{fig:fig3} clearly imply that the level density is strongly reduced for the neutron-deficient $^{115}$Te and $^{114}$Sb which are away from the stability line. For instance, by incorporating the level density parameters given by Eq.~\ref{eq:eq4} into the Fermi gas level density expression (Eq.~\ref{eq:eq1}) the estimated NLD of the neutron-deficient $^{115}$Te becomes $\approx  $10 times lower than that of the $\beta $-stable $^{115}$Sn around the neutron separation energy. \\
\indent The present experimental observation is in contrast to the high energy fusion evaporation studies of Charity {\it et al.}~\cite{Char1} and Moro {\it et al.}~\cite{Moro} performed to investigate the isospin dependence in NLD. While Charity {\it et al.}~\cite{Char1} did not found any convincing evidence for the neutron-proton asymmetry dependence of NLD, Moro {\it et al.}~\cite{Moro} showed that the $N$-$Z$ prescription provide a somewhat better explanation of their data although the results do not discard an isospin independent form of the level density parameter. 
\begin{figure}
\begin{center} 
\includegraphics[scale=0.60]{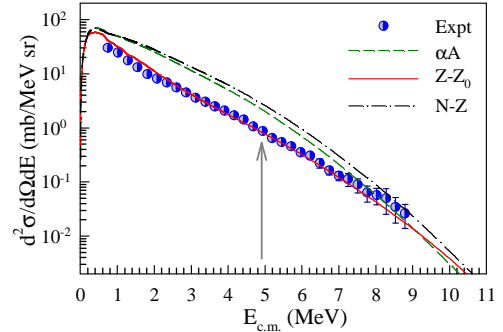}
\caption{\label{fig:fig3} The same as Fig.~\ref{fig:fig1} but for the $^4$He~+~$^{112}$Sn reaction.}
\end{center}
\end{figure}
It should be mentioned here, that an extension of the Al. Qurashi parameterizations to very high energies and high spins could be questionable. The situation at higher energies become complicated as the asymptotic level density parameter itself may show significant energy dependence~\cite{Char1,Char3}. Besides, in high energy heavy-ion fusion reactions the level density parameter gets averaged out over a large number of effective decay channels and may not correspond to one or two specific nuclei of interest. Therefore, light-ion induced low energy reactions could possibly be the most suitable probe to investigate the neutron-proton asymmetry dependence of level density.\\
\indent To investigate the observed variation of the level density parameter we have calculated the single-particle energy levels of several $A=115$ isobars around the $\beta$ stable $Z$. For this purpose, a microscopic shell-model~\cite{Garciaa} with the Wood-Saxon mean field defined using the Rost parameters~\cite{Rost,Cwiok} is used. Subsequently, the occupation probabilities of these single particle levels are calculated at different temperatures ($T$) by following the Fermi distribution function and, as proposed in Ref.~\cite{Bethe}, the corresponding excitation energies ($E$) are extracted by adding the single-particle energies of occupied levels~\cite{Senthil}. The level density parameter, defined within the Fermi gas model, is obtained by employing the Fermi gas formula $E=aT^2$. Finally, the corresponding asymptotic values of the level density parameter are extracted by using the Ignatyuk formula (Eq.~\ref{eq:eq6})~\cite{Igna}. We have used 26 harmonic oscillator shells to define the basis states, and the choice of this basis size reduces the uncertainty in the level density parameter below 5$\%$. The results have been plotted in Fig.~\ref{fig:fig4} (red dashed line) and compared with the phenomenological $Z-Z_0$ form (green continuous line) with original parametrization of Ref.~\cite{Qura2}. The shaded region in the plot indicates the theoretical uncertainty of 5$\%$ as mentioned above.\\
\begin{figure}
\begin{center}
\vspace{0.3cm}
\includegraphics[scale=0.50]{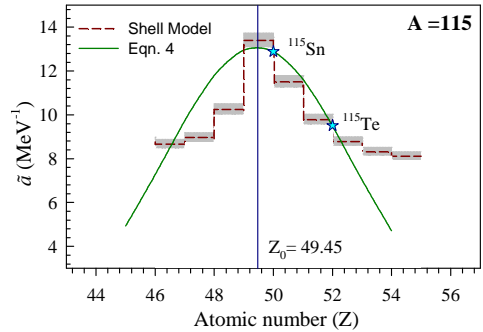}
\caption{\label{fig:fig4} The prediction of the $Z-Z_0$ form (Eq.~\ref{eq:eq4}) of the level density parameter (continuous line) is compared with a microscopic shell-model calculation (dashed line). The shaded region corresponds to the theoretical uncertainty in the shell model prediction. The nuclei investigated in the present work are indicated by the symbols.}
\end{center}
\end{figure}
Evidently, the $\tilde{a}$, calculated microscopically, shows the same trend as described by the phenomenological form given in Eq.~\ref{eq:eq4}. Particularly, both of them reasonably agree for the nuclei investigated in the present work. However, Eq.~\ref{eq:eq4} underestimates the shell-model prediction as the system departs considerably from $Z=Z_0$. It suggests, the simple empirical form of the level density parameter (Eq.~\ref{eq:eq4}) may not be sufficient to calculate level densities for  nuclei that lie far away from the valley of stability. Moreover, the shell-model $\tilde{a}$ shows a sharper fall at $Z<Z_0$ compared to the slope at $Z>Z_0$. Further experimental confirmations are required to understand this asymmetric variation along the isospin axis.\\ 
The present analysis clearly established that the level density parameter depends on $N$ and $Z$ independently rather than a simple function of $A$, and its value reduces as $N/Z$ changes from the value around the valley of stability. A significant reduction of level density for proton or neutron-rich nuclei compared to the stable ones as suggested would have a profound effect on the nucleosynthesis calculations, which typically involve ($p,\gamma $) or ($n,\gamma $) reaction channels under conditions such that successive proton or neutron captures can occur. A substantial reduction of level density will try to inhibit repeated captures which take the nucleus towards the drip line. Such a condition would change the balance between $\beta$-decay and capture and would eventually push the paths for $rp$- and $r$-process nucleosynthesis closer to the valley of stability.\\

\indent In summary, the neutron energy and angular distribution have been measured in the $p$~+~$^{115}$In, and $^4$He~+~$^{112}$Sn reactions in the compound nuclear excitation energy range of $\approx $18~-~26 MeV. Statistical model analysis of the backward angle neutron spectra was carried out to investigate the $N$, $Z$ dependence of nuclear level density parameter. It was observed that experimental data for the two reactions could be explained simultaneously by using a parametrization of the level density parameter that reduced its value as the nuclei move away from the valley of stability. Another form of $a$ which lowered the level density parameter as iso-spin is increased at fixed $A$ could not explain the data. The observed variation of the LDP around the $\beta $-stable $Z$ has been supported by a microscopic shell model calculation. Thus the present study provided a clear evidence for the reduction of nuclear level density away from the $\beta $-stability line. Further experimental data for neutron- or proton-rich nuclei in different mass regions will be useful for the systematic understanding of the $N$, $Z$ dependence of NLD. \\

\indent The authors would like to acknowledge the VECC Cyclotron operators for smooth running of the accelerator during the experiment. The authors are also thankful to Dr. Haridas Pai for providing the $^{112}$Sn target.

\end{document}